\documentclass[12pt,dvips]{article}
\usepackage{epsfig}

\usepackage{amssymb}
\usepackage{amsmath}
\usepackage{setspace}
\usepackage[numbers]{natbib}

\begin{document}

\renewcommand{\title}{Influence of aggregate size and fraction on shrinkage induced micro-cracking of mortar and concrete}

\begin{center} \textbf{\title} \end{center}

\begin{center} \underline{Peter Grassl}$^{1}$, Hong S. Wong$^{2}$ and Nick R. Buenfeld$^{2}$\\
$^{1}$~Department of Civil Engineering, University of Glasgow, Glasgow, UK; $^{2}$~Department of Civil Engineering, Imperial College London, London, UK

12th international conference on fracture (ICF-12)

Last modified: 30th October 2008

\end{center}

\textbf{Abstract}

In this paper, the influence of aggregate size and volume fraction on shrinkage induced micro-cracking and permeability of concrete and mortar was investigated.
Nonlinear finite element analyses of model concrete and mortar specimens were performed.
The aggregate diameter was varied between 2~and~16~mm. Furthermore, a range of volume fractions between 0.1 and 0.5 was studied. 
The nonlinear analyses were based on a 2D lattice approach in which aggregates were simplified as monosized cylindrical inclusions.
The analysis results were interpreted by means of crack width and change of permeability.
The results show that increasing aggregate diameter (at equal volume fraction) and decreasing volume fraction (at equal aggregate diameter) greatly increases permeability.

\textbf{1. Introduction}

Drying of cement based composites, such as concrete and mortar, induces cracking if shrinkage of the constituents is either internally or externally restrained.
For example, non-uniform drying leads to a moisture gradient, which results in non-uniform shrinkage of the specimen.
Surface regions shrink faster than the inner bulk material, which results in surface cracking \cite{BazRaf82}.
Additionally, shrinkage might be restrained by aggregates within the composite \cite{Hob74}.
Aggregate-restrained shrinkage can lead to micro-cracking, which strongly influences the transport properties of the material \cite{BisMie02, WonZobBue08}.
However, the evolution of micro-cracks and their dependence on the size and volume fraction of aggregates is not fully understood yet.
In \cite{WonZobBue08} it was observed that permeability increases with increasing aggregate size at a constant aggregate fraction. This result is surprising, since an increase of the aggregate size at a constant aggregate fraction is usually accompanied by a decrease in the volume of interfacial transition zones (ITZs), which are known to be more porous than the cement paste.
One hypothesis is that an increase of aggregate diameter at constant aggregate fraction results in an increase of micro-crack width, which is closely related to permeability \cite{WitWanIwaGal80}.
The objective of this work was to establish whether this size effect really occurs.
This will undoubtedly enhance the understanding of the link between micro-structure and macro-property, in particular the effect of microcracking on mass transport, which is a critical aspect for predicting durability and service-life.

In the present work, shrinkage induced micro-cracking of concrete was analysed by means of the nonlinear finite element method.
A lattice approach was used in combination with a damage-plasticity constitutive model, which was designed to result in mesh-independent responses \cite{GraRem08}.
This lattice approach is robust and computationally efficient. 

The present study is based on several simplifications.
Shrinkage is represented by an eigenstrain, which was uniformly applied to the cement matrix only.
This is representative of autogenous shrinkage, but does not fully represent transient non-uniform shrinkage due to moisture gradients which may lead to more cracking near the surface than in the centre of a concrete element.
Furthermore, the only inclusions considered were aggregates, which were embedded in a uniform cement paste. 
Micro-cracking due to other inclusions, e.g. unhydrated cement and calcium hydroxide crystals, was not considered.
Aggregates were assumed to be separated from the cement paste by interfacial transition zones (ITZs), which were modelled to be weaker and more brittle than the cement matrix. 
Furthermore, the study was limited to two-dimensional plane stress analyses, in which aggregates are idealised as cylindrical inclusions of constant diameter.

\textbf{2. Modelling approach}

\begin{figure}
\begin{center}
\begin{tabular}{cc}
\epsfig{file=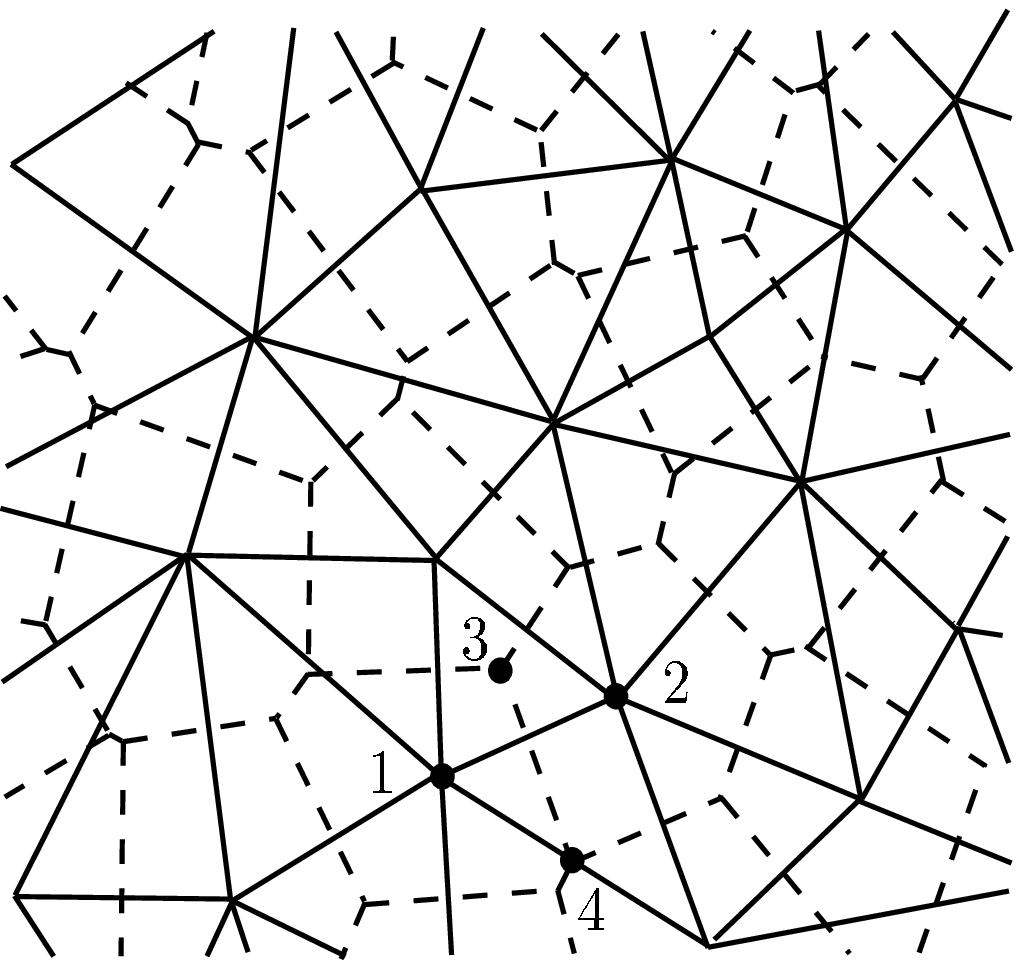,width=6cm} & \epsfig{file=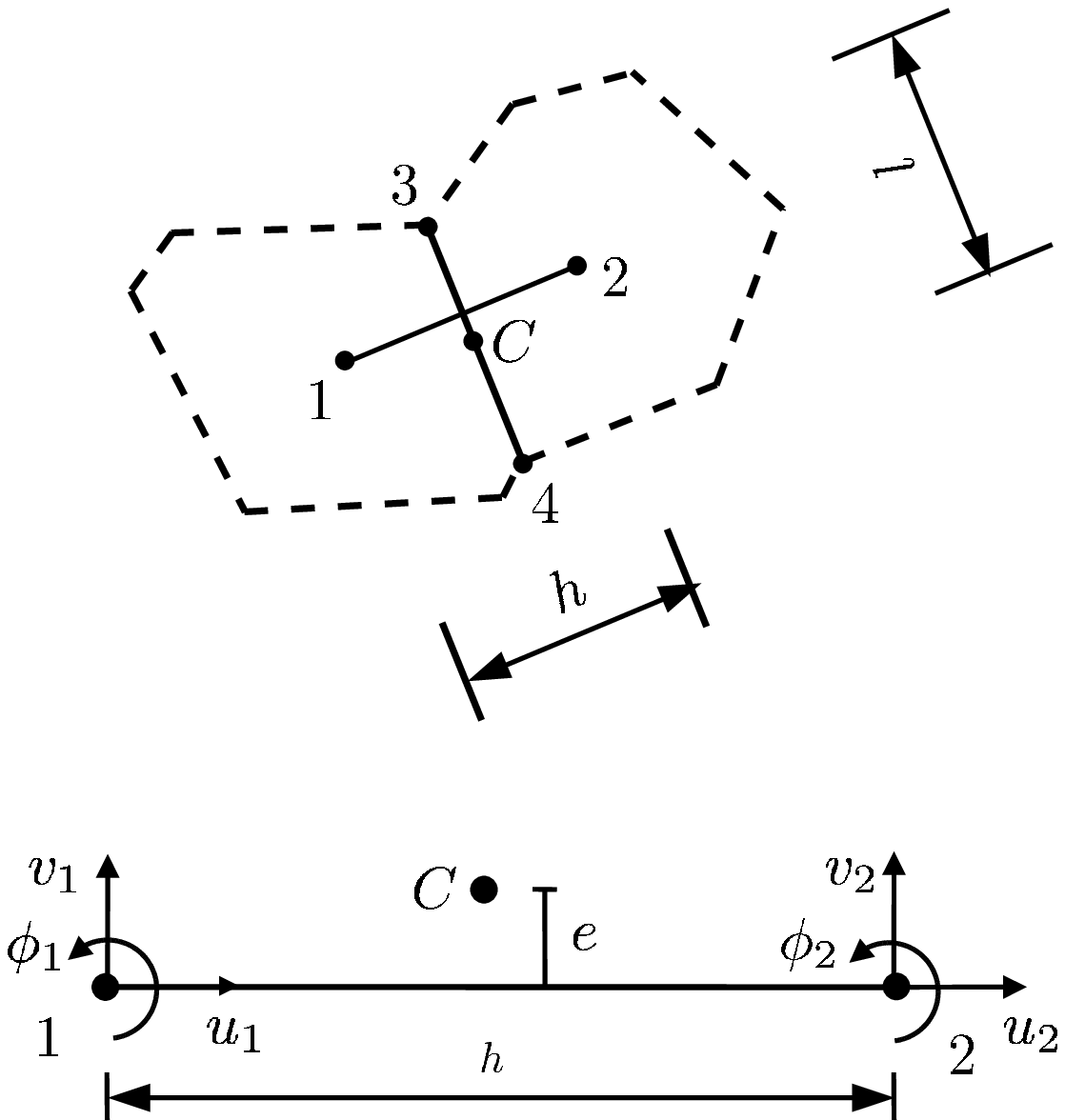,width=6cm} \\
(a) & (b)
\end{tabular}
\end{center}
\caption{Discretisation: (a) Lattice elements (solid lines) and cross-sections (dashed lines) obtained from the Delaunay triangulation and dual Voronoi tessellation, respectively. (b) Lattice element with degrees of freedom in the local co-ordinate system.}
\label{fig:vorDel}
\end{figure}

In the present work, shrinkage induced cracking was described by means of a lattice approach \cite{BolSai98} combined with a damage-plasticity constitutive law \cite{GraRem08}.
Nodes are placed randomly in the specimen constrained by a minimum distance $d_{\rm m}$, i.e. the smaller $d_{\rm m}$ is, the smaller the average element length (Fig.~\ref{fig:vorDel}a).
Based on these randomly placed nodes, the spatial arrangement of lattice elements is determined by a Voronoi tesselation.
The cross-sections of the lattice elements, which connect the nodes (Voronoi sides), are the edges of the Voronoi polygons (Fig.~\ref{fig:vorDel}a).
Each node has three degrees of freedom, two translations and one rotation, shown in the local coordinate system in Figure~\ref{fig:vorDel}b.
The degrees of freedom $\mathbf{u}_{\rm e} = \left\{u_1, v_1, \phi_1, u_2, v_2, \phi_2\right\}^{\rm T}$ of two nodes of a lattice element are related to the displacement discontinuities $\mathbf{u}_{\rm c} = \left\{u_{\rm c}, v_{\rm c}\right\}^{\rm T}$ at the mid-point $C$ of the cross-section.
The constitutive model of the lattice elements is a combination of plasticity and damage mechanics \cite{GraRem08}, which relates stresses to strains.
The strains $\boldsymbol{\varepsilon} = \left(\varepsilon_{\rm n}, \varepsilon_{\rm s}\right)^T$ are determined from the displacement jump $\mathbf{u}_{\rm c} = \left(u_{\rm n}, u_{\rm s}\right)^T$ at mid-point $C$ as
\begin{equation}
\boldsymbol{\varepsilon} = \dfrac{\mathbf{u}_{\rm c}}{h}
\end{equation}
These strains are related to the nominal stress $\boldsymbol{\sigma} = \left(\sigma_{\rm n}, \sigma_{\rm s}\right)^T$ as 
\begin{equation} \label{eq:totStressStrain}
\boldsymbol{\sigma} = \left(1-\omega \right) \mathbf{D}_{\rm e} \left(\boldsymbol{\varepsilon} - \boldsymbol{\varepsilon}_{\rm p} - \boldsymbol{\varepsilon}_{\rm s}\right) = \left(1-\omega\right) \bar{\boldsymbol{\sigma}}
\end{equation}
where $\omega$ is the damage variable, $\mathbf{D}_{\rm e}$ is the elastic stiffness,  $\boldsymbol{\varepsilon}_{\rm p} = \left(\varepsilon_{\rm pn}, \varepsilon_{\rm ps}\right)^T$ is the plastic strain, $\boldsymbol{\varepsilon}_{\rm s} = \left(\varepsilon_{\rm s}, 0\right)^T$ is the shrinkage eigenstrain and  $\bar{\boldsymbol{\sigma}} = \left(\bar{\sigma}_{\rm n}, \bar{\sigma}_{\rm s}\right)^T$ is the effective stress.

The plasticity part of the model is based on the effective stress $\bar{\boldsymbol{\sigma}}$ and consists of an elliptic yield surface, an associated flow rule, an evolution law for the hardening parameter and loading and unloading conditions.
A detailed description of the components of the plasticity model is presented in \cite{GraRem08}.
The initial yield surface of the plasticity model is determined by the tensile strength $f_{\rm t}$, the shear strength $s f_{\rm t}$ and the compressive strength $c f_{\rm t}$.
The evolution of the yield surface during hardening is controlled by the model parameter $\mu$, which is defined as the ratio of permanent and reversible inelastic displacements.
The damage part is formulated so that linear stress inelastic displacement laws for pure tension and compression are obtained, which are characterised by the fracture energies $G_{\rm ft}$ and $G_{\rm fc}$, respectively. 

The equivalent crack opening $\tilde{w}_{\rm c}$ is defined by the equivalent inelastic displacement, which for the present damage-plasticity model is defined as 
\begin{equation} \label{eq:equivCrack}
\tilde{w}_{\rm c} = \|\mathbf{w}_{\rm c}\|
\end{equation}
where
\begin{equation}
\mathbf{w}_{\rm c} = h \left(\boldsymbol{\varepsilon}_{\rm p} + \omega \left(\boldsymbol{\varepsilon} - \boldsymbol{\varepsilon}_{\rm p}\right)\right)
\end{equation}
The inelastic displacement vector $\mathbf{w}_{\rm c}$ is composed of a permanent and reversible part, defined as $h \boldsymbol{\varepsilon}_{\rm p}$ and $h \omega \left(\boldsymbol{\varepsilon} - \boldsymbol{\varepsilon}_{\rm p}\right)$, respectively.

\textbf{3. Nonlinear finite element analysis of shrinkage induced micro-cracking}

Shrinkage induced micro-cracking was analysed by means of the nonlinear lattice approach described above.
The elements representing the cement paste were subjected to an incrementally applied uniform shrinkage strain up to $\varepsilon_{\rm s} = 0.5$~$\%$ (Eq.~\ref{eq:totStressStrain}).
This value was chosen for the simulation to represent a relatively severe shrinkage of the neat cement paste on first-drying to low humidities. 
The influence of aggregate volume fraction and aggregate diameter was studied. Aggregate volume fractions $\rho = 0.5, 0.3$~and~0.1 were modelled. 
Furthermore, four different aggregate diameters ($\phi = 16, 8, 4,$~and~$2$~mm) were used.
\begin{figure}
\begin{center}
\begin{tabular}{cc}
\epsfig{file=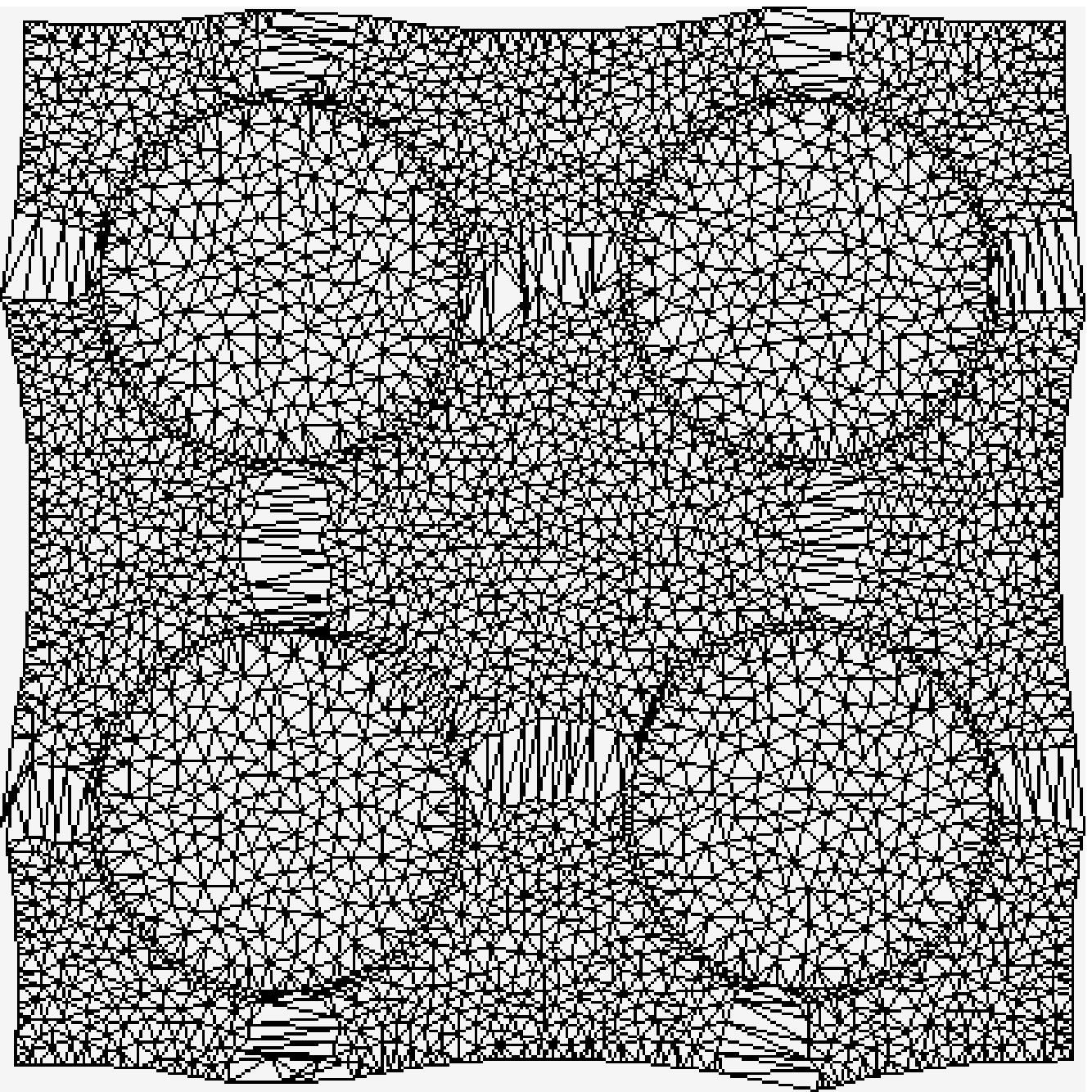, width=5cm} & \epsfig{file=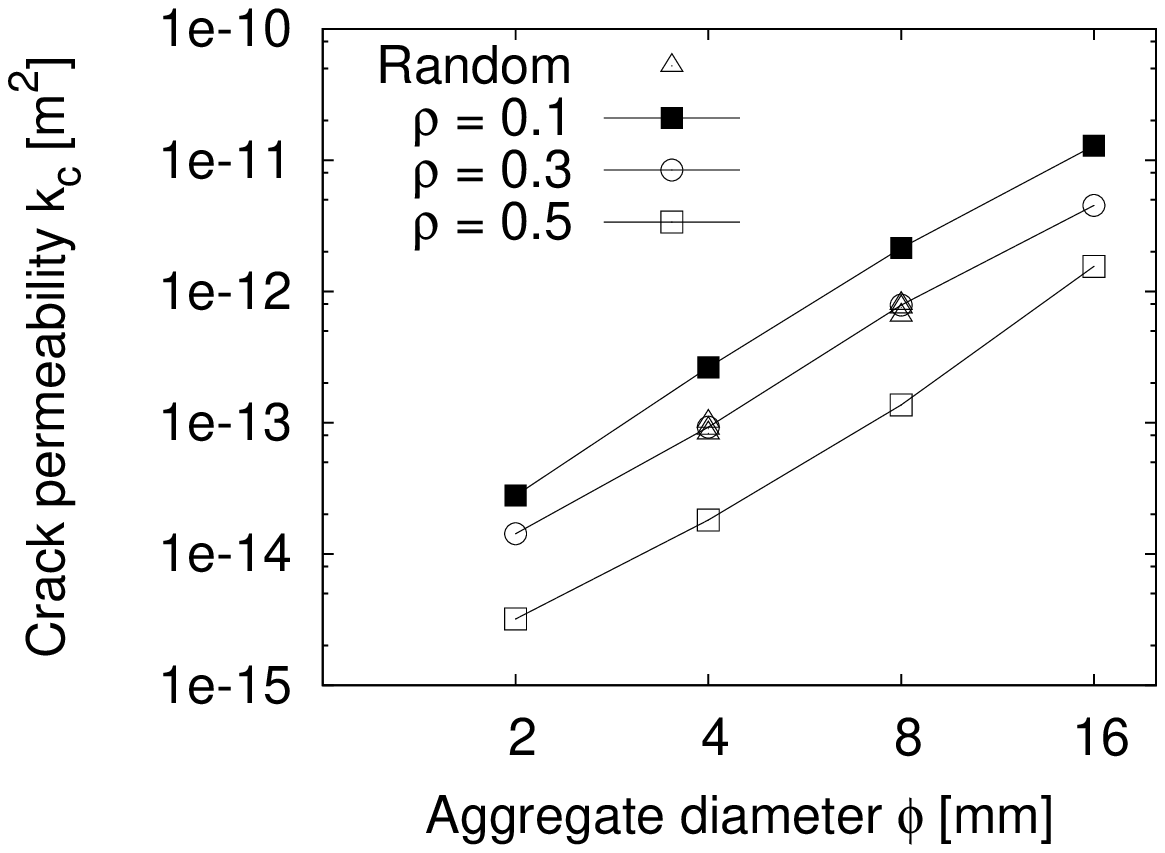, width=8cm}\\
(a) & (b)
\end{tabular}
\end{center}
\caption{Analysis results at a shrinkage strain of $\varepsilon_{\rm s} = 0.5$~$\%$: (a) Deformations (magnified by a factor of 50) for $\phi=16$~mm. (b) Crack permeability $k_{\rm c}$ (Eq.~\protect\ref{eq:crackPerm}) versus aggregate diameter $\phi$ for volume fractions of $\rho = 0.1, 0.3$~and~$0.5$.}
\label{fig:results}
\end{figure}
The length $L$ of the specimen  
\begin{equation}\label{eq:length}
L = \sqrt{\dfrac{\pi \phi^2}{\rho}}
\end{equation}
Thus, the smaller the aggregate diameter $\phi$ and the greater the volume ratio $\rho$, the smaller is the specimen length $L$ (Fig.~\ref{fig:results}).
The smallest separation distance between aggregate particles, i.e. minimum width of the cement paste decreases with increasing aggregate volume fraction at constant aggregate size, as would be expected in the case of real mortars with increasing aggregate fraction. 
At constant aggregate fraction, the separation between aggregate particles increases with increasing aggregate size, similar to the case of a mortar compared to concrete at the same aggregate content. 
Aggregates were modelled elastically. 
Lattice elements crossing the boundary between aggregates and cement paste represent the average response of the interfacial transition zones and the two adjacent material phases.
In all the analyses, the length of the lattice elements is significantly greater than the width of the highly non-uniform interfacial transition zones, which is usually in the range of 10s of $\mu$m \cite{ScrCruLau04}.
Therefore, the stiffness of these lattice elements was approximated as an average of the Young's modulus of aggregate and mortar.
The strength of these lattice elements is determined by the strength of the ITZ, which is the weakest link.
Here, the strength and fracture energy ratio of cement paste and ITZ was chosen as 2.  
This ratio is an approximate value for samples with a relatively weak ITZ. 
For an aggregate diameter of $\phi=16$~mm and a volume fraction of $\rho = 0.3$, the deformed mesh is shown in Fig.~\ref{fig:results}a.
The deformations are localised in a few cracks, which connect the aggregates in a regular square pattern. 
Qualitatively very similar crack patterns were obtained for the other aggregate diameters and volume fractions.
This pattern of cracking seems consistent to that observed by Hsu \cite{Hsu63} in '2D' model samples made of sandstone discs arranged in a square grid and filled with paste that is subsequently subjected to drying shrinkage. 
Depending on the separation between the aggregates, cracks were seen to occur at the interface (i.e. bond cracks), near the shortest distance and diagonally at the largest distance between aggregates. 

The average crack width increases with increasing aggregate size at constant aggregate volume fraction. At constant aggregate size, the average crack width increases with decreasing aggregate volume fraction.
Crack width is closely related to transport properties of concrete, in particular in the case of flow under a pressure gradient.
Assuming that the paste matrix is dense so that flow occurs predominantly through the cracks, the permeability $k_{\rm c}$ in the out-of-plane direction due to cracking can be described by the cube of the crack width as
\begin{equation} \label{eq:crackPerm}
k_{\rm c} = \dfrac{\xi}{A} \sum_{i = 1}^{n_{\rm c}} l_{i} \tilde{w}^3_{\rm{c} i}
\end{equation}
where $\xi$ is a material constant \cite{WitWanIwaGal80}.
The influence of aggregate diameter and volume fraction on $k_{\rm c}$ is shown in Fig.~\ref{fig:results}b on a log-log scale for $\xi=1$.
Note that what is of interest here is not the actual value of the estimated permeability, but the change in permeability caused by varying either the aggregate volume fraction or particle size.
At a constant aggregate volume fraction, increasing the aggregate diameter from 2 to 16~mm caused approximately a 2.5 orders of magnitude increase in permeability. At constant aggregate diameter, increasing the aggregate volume fraction from 0.1 to 0.5 produced about 1 order of magnitude decrease in permeability.
The aggregate diameter influences the permeability strongly, since the crack width, which increases with increasing aggregate diameter, enters Eq.~(\ref{eq:crackPerm}) in its cube.

\textbf{4. Conclusions}

In the present work the influence of aggregate size and volume fraction on shrinkage induced micro-cracking was studied numerically by means of the nonlinear finite element method. 
The work has shown that the average crack width increases with increasing aggregate diameter and decreasing volume fraction.
The permeability, which is related to the cube of the crack width, increases with increasing aggregate diameter and decreasing volume fraction
Thus, the aggregate diameter and volume fraction influence the formation of microcracks induced by aggregate restrained shrinkage, which is in agreement with experimental observations in \cite{WonZobBue08}.
However, the representation of aggregates as circular inclusions and the use of one aggregate size oversimplifies the micro-structure of concrete, which does not allow a direct comparison with experimental results.
In future work, the present modelling approach will be extended to 3D and the use of realistic aggregate grading curves will be included by applying multi-scale analysis approaches.

\textbf{Acknowledgements}

HSW and NRB acknowledge the financial support from the Engineering and Physical Sciences Research Council, UK. 

\bibliographystyle{plainnat}
\bibliography{general}

\clearpage

\end{document}